\documentclass[journal]{IEEEtran}

\usepackage{amsmath}
\usepackage{graphicx}
\usepackage{booktabs}  
\usepackage{fullpage}
\usepackage{times}
\usepackage{fancyhdr,graphicx,amsmath,amssymb}
\usepackage[ruled,vlined]{algorithm2e}
\include{pythonlisting}
\usepackage{amsmath, amssymb}
\usepackage{makecell}
\usepackage{multirow}
\usepackage{enumitem}
\usepackage{threeparttable}
\usepackage{xcolor}
\usepackage{colortbl}
\usepackage{float}

\begin{document}

\title{Quantum Power Flow}

\author{Fei~Feng,~\IEEEmembership{Student Member,~IEEE}, 
         Yifan~Zhou,~\IEEEmembership{Member,~IEEE} and Peng~Zhang,~\IEEEmembership{Senior Member,~IEEE} 
\thanks{This work was supported in part by the National Science Foundation under Grant No. OIA-2040599, the Advanced Grid Modeling Program (AGM) under Department of Energy's Office of Electricity (OE), and in part by the Department of Energy's Office of Electricity. This research used resources of the Oak Ridge Leadership Computing Facility, which is a DOE Office of Science User Facility supported under Contract DE-AC05-00OR22725.}
\thanks{The authors are are with the Department of Electrical and Computer Engineering, Stony Brook University, Stony Brook, NY 11794-2350, USA (e-mail: P.Zhang@stonybrook.edu).}% <-this % stops a space
}

\markboth{}%Journal of Class Files,~Vol.~14, No.~8, August~2019
{Shell \MakeLowercase{\textit{ et al.}}:  Bare Demo of IEEEtran.cls for IEEE Journals}

\maketitle

\begin{abstract}
This letter is a proof of concept for quantum power flow (QPF) algorithms which underpin various unprecedentedly efficient power system analytics exploiting quantum computing. 
Our contributions are three-fold: 
1) Establish a quantum-state-based fast decoupled model empowered by Hermitian and constant Jacobian matrices;
2) Devise an enhanced Harrow-Hassidim-Lloyd (HHL) algorithm to solve the fast decoupled QPF; 
3) Further improve the HHL efficiency by parameterizing quantum phase estimation and reciprocal rotation only at the beginning stage. 
Promising test results validate the accuracy and efficacy of QPF and demonstrate QPF's enormous potential in the era of quantum computing. 
\end{abstract}

% Note that keywords are not normally used for peerreview papers.%derivative-based approaches with inclusion of classical direct iterative algorithms for linear systems of equations scale with time $O(N^2\sim N^3)$ for $N \times N $ system, which remains an open challenge to bit-based computation technology~\cite{8309382}. 
\begin{IEEEkeywords}
quantum power flow, quantum computing, fast decoupled power flow.
\end{IEEEkeywords}
\IEEEpeerreviewmaketitle
\section{Introduction}
\IEEEPARstart{T}{raditional} tools for real-time operation of modern power systems, such as probabilistic power flow, $N$$-$$x$ security screening and Monte Carlo methods, remain to be intractable problems. Power flow equations, if solved by the classical direct iterative algorithms,  %for linear systems of equations
scale with time as $O(N)$ for an $N$$\times$$N$ system~\cite{DommelNotes}. However, tremendous amount of repetitive power flow calculations are needed to analyze the impact of uncertainties (e.g., output from distributed energy resources, fluctuating demands, and random failures or faults) through traditional methods such as probabilistic power flow, making the exiting approaches impossible to meet the real-time operation requirements~\cite{NSFproposal}.

%flow calculation is an indispensable foundation for power system planning and operations. Probability power flow, N-$x$ security assessment and various system control, protection and energy management functions  require  fast online solutions of massive power flow cases for managing today's power systems featured by highly variable resources and prosumer loads as well as increasingly frequent natural and man-made disturbances~\cite{NSFproposal}.

%However, even solving linear system for bit-based computation technology scale with time $O(N^2\sim N^3)$ for $N$$\times$$N$ system, making the massive iterative power flow calculations an intractable challenge.

%Quantum computing enables exponential speedups over classical computers, which is derived from the possibility to prepare and maintain complex superpositions of quantum states across many quantum degrees of freedom as well as to provide entanglement between the states of the system~\cite{preskill2012quantum}. An important result in recent years has been Harrow-Hassidim-Lloyd (HHL) algorithm, a quantum method for solving linear equations ~\cite{PhysRevLett.103.150502}. It effectively accelerate the analysis of sparse systems, which exactly matches the power system characteristics. Meanwhile, some hybrid algorithms such as Variational Quantum Linear Solver and quantum random walk algorithm are also developed for tradeoff between quantum depth and noise.

Theoretically, quantum computing algorithms can achieve exponential speedups over classical methods using noisy-free quantum computers~\cite{nielsen2000quantum,Chen_2019}. %An important result of quantum computing research in recent years has been the Harrow-Hassidim-Lloyd (HHL) algorithm, a quantum method for solving linear equations ~\cite{harrow2009quantum}.%In this proposal, we seek a practical path forward for science on near-term devices at current noisy-intermediate-scale quantum (NISQ) computers, while still laying the foundations (software, algorithms, workforce skills, etc.) for science on noisy-free quantum computers of a distant future (5-10 years)
This work is the first attempt of leveraging quantum supremacy to resolve the intractable challenge related to power flow calculations. The key innovation is to architect a practical quantum power flow (QPF) model and solver through an improved Harrow-Hassidim-Lloyd (HHL)~\cite{harrow2009quantum} algorithm. This letter demonstrates QPF's potential to meet the growing needs of power flow calculation and support fast and resilient power system operations. %and validate it through which is expected to speed up the rapidly growing power flow calculation needs.% The main contributions of QPF lie in : 1) a quantum-enabled fast decoupled power flow model incorporating the quantum representation of power system states; 2) a HHL-based QPF algorithm considering the sparse and Hermitian Jacobian features of fast decoupled model; 3) improved HHL efficiency by constant Jacobian to allow a single execution of quantum phase estimation and reciprocal rotation.

\section{Quantum Power Flow}
%The conclusion goes here.

%To tackle the meshed microgrid power flow, an enhanced NR method is proposed using a novel approach where hierarchical control is incorporated.
%\subsection{Defining Buses in Microgrid}QPF is established on the fast decoupled approach which derived from Newton algorithm. The main difference is its modified power flow formulations considering the unique feature of power system in which voltage angles are mainly affected by active power, voltage magnitudes are mainly affected by reactive power, respectively. Then, Jacobian matrix can be simplifed by decoupling reactive power with voltage angle and active power with voltage magnitude. Eventually, the modified formulations can be devised as follows,

\subsection{Fast Decoupled QPF Formulation}
%This letter devises a fast decoupled QPF formulation. 
Fast decoupled power flow~\cite{stott1974fast} is a most widely used variant of Newton-Raphson power flow owing to its excellent computational efficiency and convergence performance.
It adopts constant Jacobian matrices based on the fact that in a bulk power grid voltage angles are mainly related to active power and voltage magnitudes to reactive power, and thus reduces the costs for updating the Jacobian matrix in each iteration.
Inspired by the fast decoupled approach, %\eqref{E1} and \eqref{E2} 
we extend the traditional power flow into a quantum computing model: 
\begin{align}
 & \lvert \mathbf V^{-1} \Delta\mathbf P \rangle = \mathbf B^{'}
 \lvert \mathbf V \Delta\boldsymbol{\theta}   \label{E1} \rangle\\
 &  \lvert \mathbf V^{-1} \Delta\mathbf Q \rangle = \mathbf B^{''} 
 \lvert \Delta \mathbf V \rangle \label{E2}
\end{align}
\noindent where $\lvert \cdot \rangle$ denotes the normalised quantum states, which will be futher explained in Subsection~\ref{sec:HHL}; $\Delta \mathbf V$ and $\Delta\boldsymbol{\theta} $ are the differences of voltage magnitudes and angles, respectively;  $\mathbf B^{'}$ and $\mathbf B^{''} $ are coefficient matrices derived from the admittance matrix. Given $\Delta \mathbf V$ and $\Delta\boldsymbol{\theta} $, power mismatches $\Delta\mathbf {S}=[\Delta\mathbf{{P}} ,\Delta\mathbf{Q}]^T$ can be updated by 
\begin{equation}\label{E3}
\begin{aligned}
  \Delta\mathbf S&=\begin{bmatrix}
\mathbf {S}-\mathbf{\Bar{Y}(\boldsymbol{\theta})} \cdot \mathbf{V} \circ \mathbf{V}  
\end{bmatrix}
   \end{aligned}
  \end{equation}

\noindent where $\mathbf {S}=[\mathbf{{P}} ,\mathbf{Q}]^T$ represents the active/reactive power injections, $ \mathbf{\Bar{Y}(\boldsymbol{\theta})}$ is the  admittance matrix, $\circ$ means Hadamard product. 
\vspace{-10pt}
\subsection{HHL-based QPF Algorithm } \label{sec:HHL}
Apart from having constant Jacobian matrices, a striking feature of the QPF model is that %in (1)-(2) (i.e., 
$\mathbf B^{'}$ and $\mathbf B^{''} $ are both Hermitian and sparse. This allows for a direct translation of the classical power flow into the quantum language. %perfectly arouses the efficacy of most quantum algorithms in dealing with Hermitian matrices. 

%To exploit the quantum computing capability, the classical matrices in (1)-(2) should be encoded into the quantum language. 
Taking (2) as an example, the spectral decomposition of $\mathbf B^{''} $ can be devised as
  \begin{equation}\label{E4}
   \begin{aligned}
 \mathbf B^{''}=\sum_{i=1}^\zeta {  \lambda_i\lvert b^{''}_i\rangle\langle b^{''}_i\lvert}
   \end{aligned}
  \end{equation}  %\vspace{-10pt}
\noindent where $\lambda_i$ and $\lvert b^{''}_i\rangle$ are the $i^{th}$ eigenvalue and eigenvector of $\mathbf B^{''}$. Written in the eigenbasis of $\mathbf B^{''}$, $\lvert \mathbf V^{-1} \Delta\mathbf Q\rangle=\sum_{i=1}^\zeta { \alpha_i \lvert b^{''}_i\rangle}$, which gives%. Therefore, the differences of voltage magnitudes $\lvert \Delta \mathbf V\rangle$ can be devised as
\begin{equation}\label{E5}
   \begin{aligned}
 \lvert \Delta \mathbf V\rangle=\mathbf B^{''-1}\lvert \mathbf V^{-1} \Delta\mathbf Q\rangle=\sum_{i=1}^\zeta {  \lambda_i^{-1}\alpha_i \lvert b^{''}_i\rangle}
   \end{aligned}
  \end{equation}  %\vspace{-10pt} of the form $\lvert \Delta\boldsymbol{\theta}\rangle$ and $\lvert \Delta \mathbf V\rangle$

An improved HHL algorithm, as shown in Fig.1, is developed to achieve the aforementioned QPF computations in three steps. Three registers ($R_c,R_v,R_l$) are initialized at the beginning of QPF. %{\color{blue}Quantum operators act on vectors to achieve target quantum states.} % of circuit. 
$R_c$ contains the binary representation of the eigenvalues of $\mathbf B^{'}$ and $\mathbf B^{''}$. $R_v$ stores the qubit representation of $\lvert \mathbf V^{-1} \Delta\mathbf Q\rangle$ and $\lvert \mathbf V^{-1} \Delta\mathbf P\rangle$, and $R_l$ regulates the angles of operators in ancilla quantum encoding (AQE)~\cite{nielsen2000quantum}.

\subsubsection*  {Step 1: Quantum phase estimation (QPE)} It aims to determine the eigenvalue of unitary operators by using phase kickback and quantum inverse Fourier transform (QFT$^\dagger$). In the phase kickback, $R_c$ is set to $\lvert 0\rangle$ and followed by Hadamard gates to provide superposition states~\cite{harrow2009quantum}. { If register $R_c$ is $\lvert 0\rangle$, the controlled unitary operator does nothing to register $R_v$; if register $R_c$ is $\lvert 1\rangle$, then the eigenvalues of controlled unitary operators can be kicked into $\lvert 1\rangle$ on register $R_c$. }   Eventually, quantum phase estimation can pick up the binary decimals of eigenvalue $\lvert \lambda_i\rangle$. After QPE, a quantum state can be generated in the eigenbasis of $\mathbf B^{''}$ as $\sum_{i=1}^\zeta {\alpha_i\lvert \lambda_i\rangle \otimes\lvert b^{''}_i\rangle}$.

\subsubsection*  {Step 2: Inverse rotation} This step aims to kick the reciprocal of $\lambda_i$ into state $\lvert 1\rangle$ for measurement. { The reciprocal of eigenvalues from QPE can be achieved through the controlled operators in AQE} %Each eigenvalue of $\mathbf B^{''}$ from QPE is rotated by 
%regulating $Rl$
: $\lvert 0\rangle\to \sqrt{1-\frac{C^2}{\lambda_i^2}}\lvert 0\rangle+\frac{C}{\lambda_i}\lvert 1\rangle$. 
Benefiting from invariant $\mathbf B^{'}$ and $\mathbf B^{''}$, the QPE and inverse rotation are parameterized only in the first iteration to update phase angle operators, significantly improving the efficiency of HHL. 

\subsubsection*  {Step 3: Inverse QPE (QPE$^\dagger$)} {The inverse QPE subroutine disentangles the register $R_c$ to  $\lvert 0\rangle$ by using controlled operator and leaves the remaining state as:} 
  \begin{equation}\label{E6}
   \begin{aligned}
\sum_{i=1}^\zeta {\alpha_i\lvert 0\rangle \otimes\lvert b^{''}_i\rangle(\sqrt{1-\frac{C^2}{\lambda_i^2}}\lvert 0\rangle+\frac{C}{\lambda_i}\lvert 1\rangle)}
   \end{aligned}
  \end{equation}  

Once the measurement of $R_l$ is $\lvert 1\rangle$, the corresponding iterative results $\lvert \mathbf V \Delta\boldsymbol{\theta} \rangle$ and $\lvert \Delta \mathbf V\rangle$ are in post-measurement states. Then, power flow variables $\boldsymbol{\theta}$,$\mathbf{V}$ can be updated for the next iteration. The QPF iterations continues until the mismatches $\Delta\mathbf P$ and $\Delta\mathbf Q$ achieve a convergence tolerance of $\xi$. 
 
%Then, the error of variables is evaluated. 
    \begin{figure}[t!]
  \centering
  \includegraphics[width=0.49\textwidth, height=0.19\textwidth]{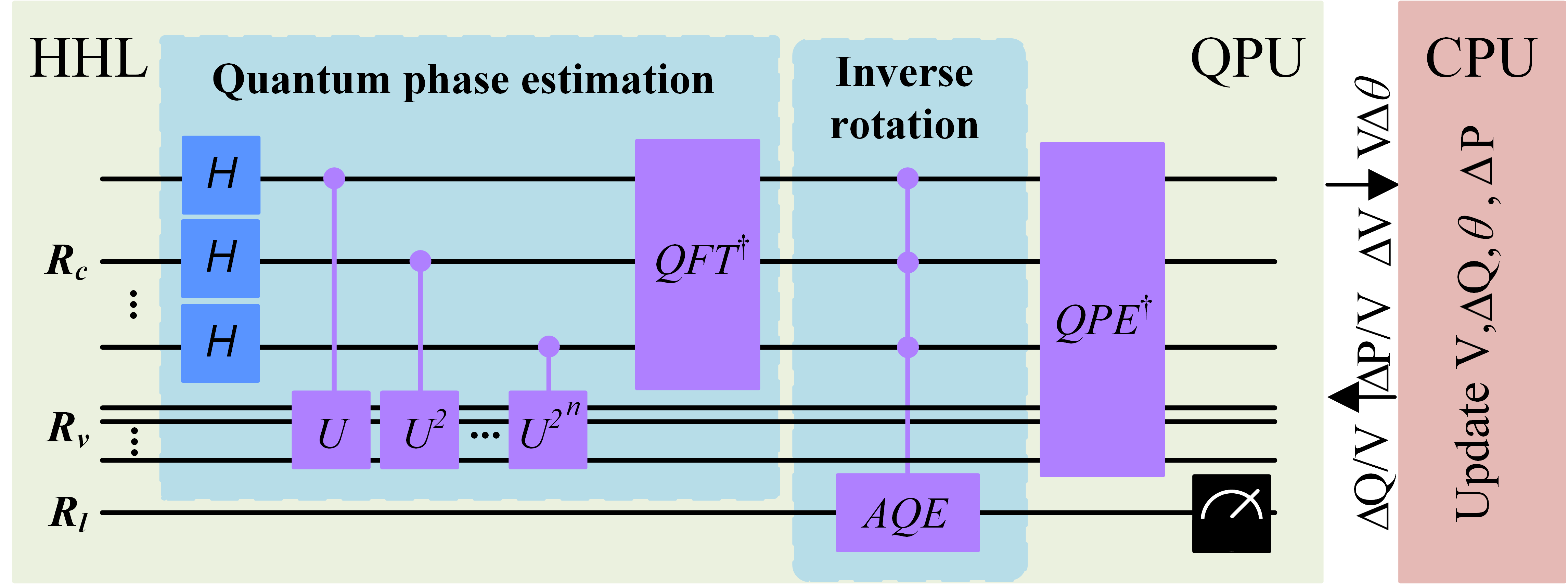}
  \caption{{ Quantum circuit architecture for the HHL-based QPF algorithm}}
  \label{F1}
  \vspace{-10pt}
\end{figure}
QPF for the first time architect an AC power flow solution in quantum computers. \textbf{Algorithm 1} presents the pseudo code of QPF. %, which does not rely on DC power flow assumptions such as  
%The superiority of fast decoupled QPF is two-fold:
%1) It pioneers a quantum-enabled power flow formulation, which bridges the unprecedentedly powerful computing capability of the quantum computers to power system analytics; 
%2) It employs the fast decoupled power flow nature, which conquers the drawbacks of non-Hermitian and inconstant Jacobian matrices in Newton-type power flows, and defeats the flaws of network $r/x$ ratio assumption of the DC-type power flows (i.e., $r/x  \ll 1$).

\vspace{-6pt}
\begin{algorithm}%[H]
\SetAlgoLined
  \textbf{{In}itialize:} $\boldsymbol{\theta}$, $\mathbf{V}$, $\mathbf B^{'}$, $\mathbf B^{''}$, $\mathbf P$, $\mathbf Q$, $\xi$\;
 \While{$\Delta\mathbf P$, $\Delta\mathbf Q$\(\geq\)$\xi $ }
 {
   Update: $\Delta\mathbf P$, $\Delta\mathbf Q$ \textbf{Eq.}~(\ref{E3})\;
  \If{$1^{st}$ iteration}{
   Input: $\mathbf B^{'}$, $\mathbf B^{''}\Rightarrow$ HHL\;
   }
  Input: $\Delta\mathbf P$, $\mathbf{V}\Rightarrow$HHL $\Rightarrow\Delta\boldsymbol{\theta}$\;
  Input: $\Delta\mathbf Q$, $\mathbf{V}\Rightarrow$HHL $\Rightarrow\Delta\mathbf{V}$\;
  Update: $\boldsymbol{\theta}$, $\mathbf{V}$\;
 }
 \KwResult{$\boldsymbol{\theta}$, $\mathbf{V}$ and the branch power flow. }
 \caption{QPF Algorithm}
 \label{EMPF}
\end{algorithm}
\vspace{-10pt}

\section{Case Study}
\begin{figure}[ht]
  \centering
  \includegraphics[width=0.45\textwidth, height=0.15\textwidth]{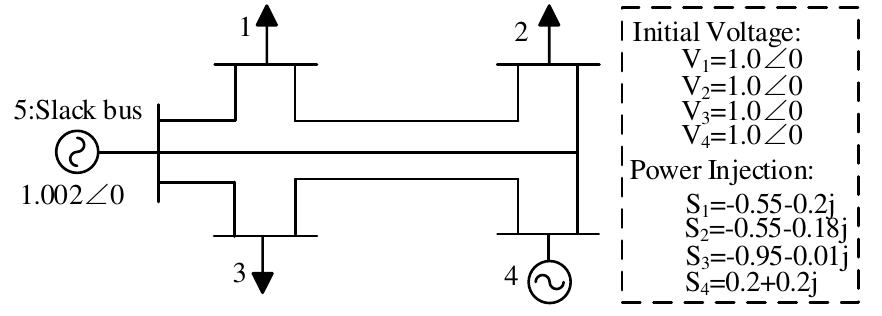}
  \caption{Five-bus system for QPF tests}
  \label{F2}
 % \vspace{-6pt}
\end{figure}

%The conclusion goes here.
The effectiveness and efficiency of QPF are verified on a five-bus test system (see Fig.~\ref{F2}). { Test I/II verifies the QPF performance on normal and stressed conditions. QPF is implemented in IBM's Qiskit (0.23.4) where the number of $R_c$ is set to 4.}

% Bus 5 is the slack bus with a specified voltage $1.002\angle 0$. Other buses are PQ buses with initial voltages $1.0\angle 0$. The active power injections at bus \{1,2,3,4\} are \{-0.55,-0.55,-0.95,0.2\}$p.u.$ , the corresponding reactive power injections are \{-0.2,-0.18,-0.01,0.2\}$p.u.$.  
\vspace{-6pt}

  \begin{figure}[ht]
  \centering
  \includegraphics[width=0.42\textwidth, height=0.22\textwidth]{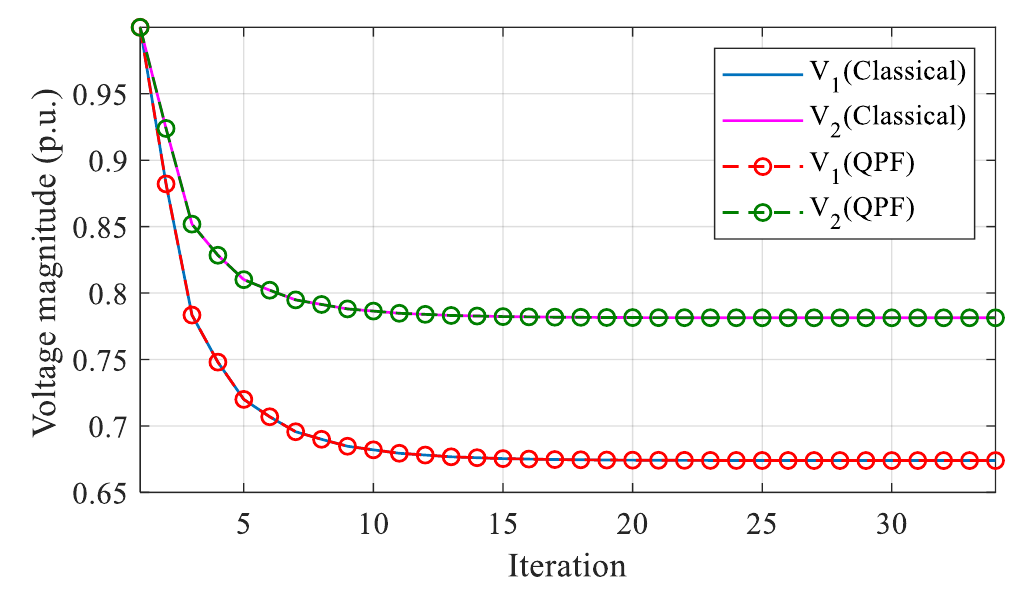}
  \caption{Voltage profiles in Test II with different methods}
  \label{F3}
 % \vspace{-6pt}
\end{figure} 
\vspace{-6pt}

\subsection{Validity of QPF}
This subsection verifies the correctness and convergence performance of QPF by comparing QPF results against those from classical fast decoupled power flow and Newton-Raphson's method.  
Table I presents the iteration process and the final power flow results under Test I. { Fig.3 shows the voltage convergence performance in Test II, where the load on bus 1 is increased to 2.2+0.8j p.u..} The following insights can be obtained: 
\begin{itemize}[leftmargin=*]
\item { The QPF results are identical to the classical results, which validates the correctness and generality of QPF.}
\item The computation process shows that HHL exhibits satisfactory accuracy at each iteration. The reason is that sufficient quantum registers are employed for quantum eigenanalysis. In this case, if the number of quantum registers is lower than 4, QPF might fail to pick up accurate results.
\item { Compared with results under the normal condition, the iteration number in Test II is increased to 34. This is because the power system is close to solvability region boundary and its voltage profiles deteriorate.}
\item QPF inherits the convergence characteristics of the classical fast decoupled method, which is slightly weaker than that of Newton's. This is because constant coefficient matrix can not adjust the calculation direction of QPF at each iteration. However, since the same convergence criteria is adopted for different power flow algorithms, the final power flow result of QPF is always as accurate as Newton's.
\end{itemize}

\begin{table}
  \caption{Voltage profiles in Test I with different methods ({\MakeLowercase{p.u.} })}\label{TableDER}
  \vspace{-5pt}
  \centering
  \begin{threeparttable}
\begin{tabular}{p{14mm}<{\centering} p{6mm}<{\raggedright} p{8mm}<{\centering} p{8mm}<{\centering} p{9mm}<{\centering} p{9mm}<{\centering}}
\toprule%\hline
 %DER bus &\multicolumn{4}{l}{ Active power } &\multicolumn{4}{l}{ Reactive power }\\
 %\multirow{2}{*}{Bus No.} &
  %\hline
   Algorithm & Iteration & $\mathbf V_3$ & $\mathbf V_4$ & $\boldsymbol{\theta}_3$ & $\boldsymbol{\theta}_4$\\
  \midrule %\hline
 \multirow{6}*{QPF} & ~~~1  & 1.0141 & 1.0282 &-0.1143 & -0.0368 \\
   &~~~2  & 0.9946 & 1.0181 & -0.1139 & -0.0340\\
  & ~~~3 & 0.9950 & 1.0183 & -0.1144 & -0.0393 \\
   &~~~4 & 0.9948 & 1.0182 & -0.1144 & -0.0393 \\
  & ~~~5 & 0.9948 & 1.0182 & -0.1144 & -0.0393 \\
  & \cellcolor{green!25}~~~6* & \cellcolor{green!25}0.9948 & \cellcolor{green!25}1.0182 & \cellcolor{green!25}-0.1144 & \cellcolor{green!25}-0.0393 \\
  \midrule% \hline
  \multirow{6}*{\makecell{Classical Fast \\ Decoupled}}& ~~~1  & 1.0141 & 1.0282 &-0.1143 & -0.0368 \\
   &~~~2  & 0.9946 & 1.0181 & -0.1139 & -0.0340\\
  & ~~~3 & 0.9950 & 1.0183 & -0.1144 & -0.0393 \\
   & ~~~4 & 0.9948 & 1.0182 & -0.1144 & -0.0393 \\
  & ~~~5 & 0.9948 & 1.0182 & -0.1144 & -0.0393 \\
  & \cellcolor{green!25}~~~6* & \cellcolor{green!25}0.9948 & \cellcolor{green!25}1.0182 & \cellcolor{green!25}-0.1144 & \cellcolor{green!25}-0.0393 \\
  \midrule% \hline
  \multirow{3}*{\makecell{Classical Newton \\ Raphson}}& ~~~1  & 1.0092 & 1.0251 &-0.1136 & -0.0382 \\
   &~~~2  & 0.9951 & 1.0183 & -0.1144 & -0.0392\\
  & \cellcolor{green!25}~~~3* & \cellcolor{green!25}0.9948 & \cellcolor{green!25}1.0182 & \cellcolor{green!25}-0.1144 & \cellcolor{green!25}-0.0392 \\
  \bottomrule%\hline
\end{tabular}
\begin{tablenotes}
\item[*]     Final power flow result
\end{tablenotes}
\end{threeparttable}
%\vspace{-7pt}
\end{table}

\subsection{QPF-based Stochastic Power Flow Analysis}
  %\vspace{-6pt}
 % \begin{figure}[ht]
 % \centering
 % \includegraphics[width=0.45\textwidth]{Ur.pdf}
 % \vspace{-0.2cm}
 % \caption{Probability density distribution of voltage magnitude  at  bus 3
%  }
 % \label{F3}
%\end{figure}
  %\vspace{-6pt}
  %\vspace{-6pt}
  %\begin{figure}[ht]
 % \centering
 % \includegraphics[width=0.45\textwidth]{Ug.pdf}
 %\vspace{-0.2cm}
 % \caption{Probability density distribution of  voltage angle at bus 3%{\color{red} Change legends to EMPF\_EP, ...}
 % }
 % \label{F4}
%\end{figure} 
  \begin{figure}[ht]
  \centering
  \includegraphics[width=0.5\textwidth]{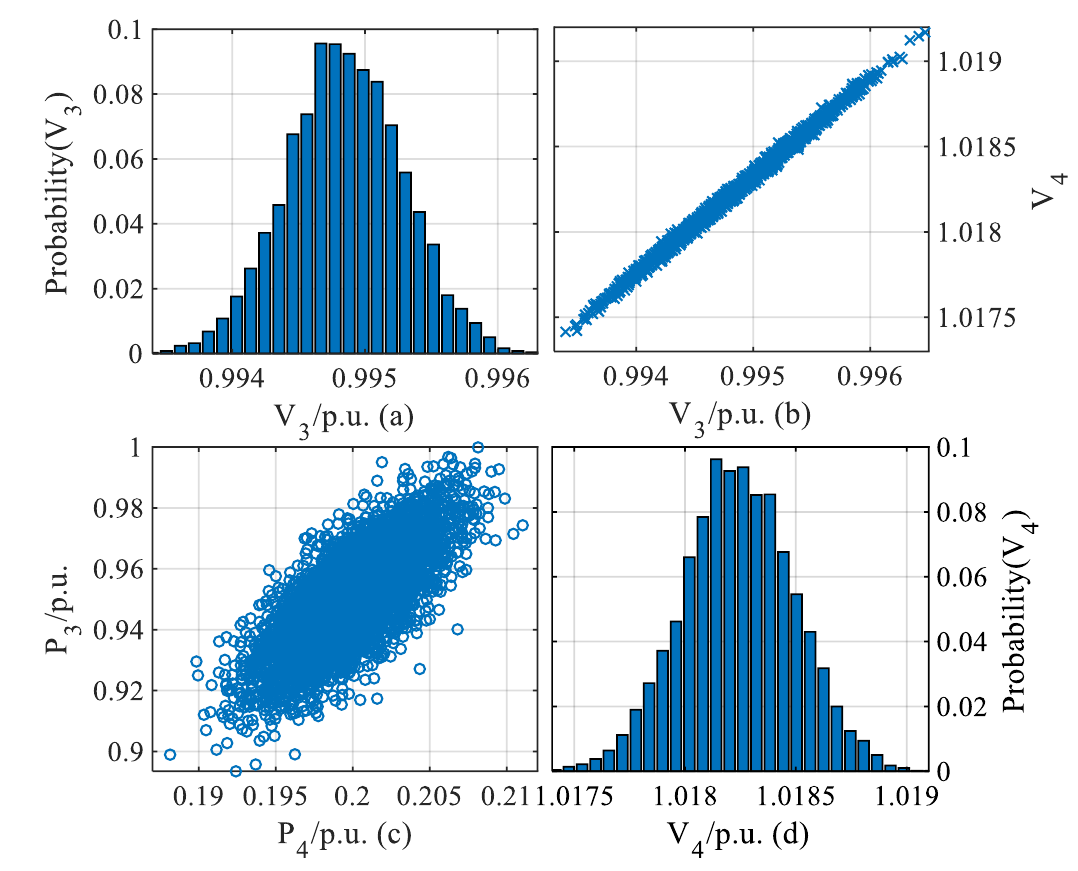}
 \vspace{-0.3cm}
  \caption{Probabilistic voltages at buses 3 and 4 and their correlations {(a) Probabilistic voltages at bus 3 (b) The correlation of voltages between buses 3 and 4 (c) The correlation of power injections between buses 3 and 4 (d) Probabilistic voltages at bus 4.}
  }
  \label{F4}
\end{figure} 
This subsection extends QPF to the stochastic power flow analysis considering correlations between system variables. % to exploit its computational potential. 
The power injections at buses 3 and 4 follow Gaussian probability distributions, where the correlation coefficient is set to be 0.75. The convergence tolerance is $\xi=10^{-5}$. 

Five thousand samples are generated randomly via Monte Carlo sampling. The probability distributions of voltage magnitudes at buses 3 and 4 as well as their correlations are obtained from QPF, as shown in Fig.~3.  
\begin{itemize}[leftmargin=*]
\item QPF is promising to be employed for stochastic power flow analyses. For instance, it can be seen that the voltage magnitude of bus 3 follows Gaussian distribution. Various correlation and dependence (i.e. Copulas, Pearson analysis) models can be readily integrated into QPF to obtain precisely the probability distributions of system states. Therefore, in the future QPF can serve as a potent tool for probabilistic system analyses. It opens a door to quantum-enabled, unprecedentedly efficient risk assessment and reliability analysis for electric grids.
\item QPF will show unprecedented computational efficiency in repetitive power flow calculations. The time complexity of classical power flow algorithm at each iteration (i.e., solving linear equations on classical computers) is $O(N)$, while QPF acquires an exponential speedup resulting in $O(log(N))$. This supremacy will be more striking for ultra-scale power systems and high-dimensional uncertainties.
\end{itemize}

QPF is still under theoretical development as it still encounters excessively large depth of quantum circuit and short coherence time in today's noisy-intermediate-scale quantum (NISQ) computers. Nevertheless, QPF lays the foundation for power flow analysis on noisy-free quantum computers of a distant future.

\section{Conclusion}
This letter opens the door for power system quantum analytics by developing a QPF algorithm. A fast decoupled QPF model is devised and solved by an enhanced HHL algorithm.
QPF is a general approach for arbitrary AC power systems, and the proof-of-concept on a small test system has been successful. Despite existing gaps for practical applications of QPF in system operations and planning due to large quantum depths, short coherence times and noises on today's quantum computers, QPF lays solid foundation for power system analytics on the next generation quantum computers with much lower noises and computational power. It is expected to grow into a enormously useful tool for energy management and security analysis. 

\bibliographystyle{IEEEtran}
\bibliography{ref}

\begin{thebibliography}{1}
\providecommand{\url}[1]{#1}
\csname url@samestyle\endcsname
\providecommand{\newblock}{\relax}
\providecommand{\bibinfo}[2]{#2}
\providecommand{\BIBentrySTDinterwordspacing}{\spaceskip=0pt\relax}
\providecommand{\BIBentryALTinterwordstretchfactor}{4}
\providecommand{\BIBentryALTinterwordspacing}{\spaceskip=\fontdimen2\font plus
\BIBentryALTinterwordstretchfactor\fontdimen3\font minus
  \fontdimen4\font\relax}
\providecommand{\BIBforeignlanguage}[2]{{%
\expandafter\ifx\csname l@#1\endcsname\relax
\typeout{** WARNING: IEEEtran.bst: No hyphenation pattern has been}%
\typeout{** loaded for the language `#1'. Using the pattern for}%
\typeout{** the default language instead.}%
\else
\language=\csname l@#1\endcsname
\fi
#2}}
\providecommand{\BIBdecl}{\relax}
\BIBdecl

\bibitem{DommelNotes}
H.~W. Dommel, \emph{Notes on Power System Analysis}, The University of British
  Columbia, 1975.

\bibitem{NSFproposal}
P.~Zhang, W.~Krawec, J.~Liu, and P.~Krstic, \emph{ASCENT: Quantum Grid:
  Empowering a Resilient and Secure Power Grid through Quantum Engineering},
  Proposal\# 2023915, National Science Foundation, Feb. 2020.

\bibitem{nielsen2000quantum}
M.~A. Nielsen and I.~L. Chuang, \emph{Quantum Computation and Quantum
  Information}.\hskip 1em plus 0.5em minus 0.4em\relax Cambridge University
  Press, 2011.

\bibitem{Chen_2019}
C.-C. Chen, S.-Y. Shiau, M.-F. Wu, and Y.-R. Wu, ``Hybrid classical-quantum
  linear solver using noisy intermediate-scale quantum machines,''
  \emph{Scientific Reports}, vol.~9, no.~1, Nov 2019.

\bibitem{harrow2009quantum}
A.~W. Harrow, A.~Hassidim, and S.~Lloyd, ``Quantum algorithm for linear systems
  of equations,'' \emph{Physical Review Letters}, vol. 103, no.~15, p. 150502,
  2009.

\bibitem{stott1974fast}
B.~Stott and O.~Alsac, ``Fast decoupled load flow,'' \emph{IEEE Transactions on
  Power Apparatus and Systems}, no.~3, pp. 859--869, 1974.

\end{thebibliography}

\end{document}